\documentclass{ametsoc}
\usepackage{xfrac}
\usepackage{amsmath}
\usepackage{amssymb}
\usepackage{chngcntr}
\usepackage{enumerate}
\usepackage{mathtools}
\usepackage{threeparttable}

\journal{jpo}

\bibpunct{(}{)}{;}{a}{}{,}

\title{On instability and mixing on the UK Continental Shelf}

\authors{Zhiyu Liu\correspondingauthor{Zhiyu Liu, State Key Laboratory of Marine Environmental Science, Xiamen University, Xiamen 361102, China.}}

\affiliation{State Key Laboratory of Marine
Environmental Science, and Department of
Physical Oceanography, College of Ocean \&
Earth Sciences, Xiamen University, Xiamen,
China}

\email{zyliu@xmu.edu.cn}

\abstract{The stability of stratified flows at locations in the Clyde, Irish and Celtic Seas on the UK Continental Shelf is examined. Flows are averaged over periods of 12--30 min in each hour, corresponding to the times taken to obtain reliable estimates of the rate of dissipation of turbulent kinetic energy per unit mass, $\varepsilon$. The Taylor--Goldstein equation is solved to find the maximum growth rate of small disturbances to these averaged flows, and the critical gradient Richardson number, $\textrm{Ri}_\textrm{c}$. The proportion of unstable periods where the minimum gradient Richardson number, $\textrm{Ri}_{\textrm{min}}$, is less than $\textrm{Ri}_\textrm{c}$ is about 35\%. Cases are found in which $\textrm{Ri}_\textrm{c} <$ 0.25; 37\% of the flows with $\textrm{Ri}_{\textrm{min}}<$ 0.25 are stable, and $\textrm{Ri}_\textrm{c}<$ 0.24 in 68\% of the periods where $\textrm{Ri}_{\textrm{min}}<$ 0.25. Marginal conditions with 0.8 $<$ Ri$_{\textrm{min}}$$/$Ri$_\textrm{c}$ $<$ 1.2 occur in 30\% of the periods examined. The mean dissipation rate at the level where the fastest growing disturbance has its maximum amplitude is examined to assess whether the turbulence there is isotropic and how it relates to the Wave--Turbulence boundary. It is concluded that there is a background level of dissipation that is augmented by instability; instability of the averaged flow does not account for all the turbulence observed in mid-water. The effects of a horizontal separation of the measurements of shear and buoyancy are considered. The available data do not support the hypothesis that the turbulent flows observed on the UK shelf adjust rapidly to conditions that are close to being marginal, or that flows in a particular location and period of time in one sea have stability characteristics that are very similar to those in another.}

\begin{document}

\maketitle

%








\section{Introduction}
Turbulent mixing on the continental shelf is a significant component of global tidal dissipation and contributes to the process known as the `continental shelf pump' related to the oceanic storage of carbon dioxide \citep{Rippeth05a}. Turbulence is generated by wind, buoyancy flux and surface waves at the sea surface, by tidal stress at the seabed and, in mid-water, by instability resulting mainly from shear. The latter is the subject of this investigation.

In a recent study, \citet{Liu10} (hereafter referred to as L10) examines the stability of baroclinic tidal flow in the Clyde Sea and its relation to dissipation and mixing. Hourly averaged density and velocity data are incorporated into the Taylor--Goldstein (T--G) equation to find the structure and rate of growth of the fastest growing small disturbances. The mean velocity, $U(z)$, is then scaled by a factor ($1+\Phi$), where $\Phi$ is a non-dimensional parameter, and the T--G equation is solved with successively decreasing values of $\Phi$ to find the conditions, at a value $\Phi=\Phi_\textrm{c}$, in which the maximum disturbance growth rate is zero. We define $\textrm{Ri}_{\textrm{min}}$ as the smallest gradient Richardson number, $\textrm{Ri}=N^2/S^2$, of the observed (un-scaled) flow, where $N(z)$ is the mean buoyancy frequency and $S = dU/dz$. The critical gradient Richardson number, $\textrm{Ri}_{\textrm{c}}$, is then $\textrm{Ri}_{\textrm{c}}$ = Ri$_{\textrm{min}}/(1+\Phi_\textrm{c})^2$. The critical gradient Richardson number determined in this way is sometimes close to the often-assumed critical value of $\sfrac{1}{4}$, but not always; the smallest value found is 0.06. A measure of whether or not the flow is in a state of ‘marginal stability’ is determined from the proximity of $\textrm{Ri}_{\textrm{min}}/\textrm{Ri}_c$ to 1 or of $\Phi_{\textrm{c}}$ to 0.

The analysis of L10 involves some pragmatic compromises that are all tested in that paper. For example, interpolation onto a 0.2 m vertical grid is found necessary in solving the T--G equation using the matrix method \citep{Monserrat96}. A limit is set for the range of unstable wavelengths, to reduce computer time. The upper and lower parts of the water column affected by bed-- or surface--generated turbulence are excluded from the analysis.

Here we apply the same analysis to examine the stability of flows in two other seas on the UK Continental Shelf, each strongly affected by the tides. This allows us to have a more general understanding on characteristics of shear instability and its role in generating turbulence in shelf seas. It is found that instability of the averaged flow does not account for all the turbulence observed in mid-water; processes possibly vortical modes or internal waves of relatively small period provide a background level of turbulence augmented by flow instability. We also test (and then refute) the hypothesis that the turbulent flows observed on the UK shelf adjust rapidly to conditions that are close to being marginal. In Section 2 data from the three seas, the Clyde (that already examined by L10), Irish and Celtic, are briefly described. Data analysis and the results are described in Section 3. This includes a general discussion of `marginal stability', which takes into account the study by \citet{DAsaro00} of the existence of a transition at the `wave--turbulence boundary' from a flow in which mixing is dominated by interactions between internal waves to one dominated by turbulence. The main conclusions are summarized in Section 4.

\section{The data}
Some details of the sites in the three seas are given in Table 1. At each site velocity profiles are obtained by a bottom-mounted acoustic Doppler current profiler (ADCP), and the density and rate of dissipation of turbulent kinetic energy per unit mass, $\varepsilon$, are measured in bursts of 5--8 Fast Light Yo-yo (FLY) profiles made every hour at positions about 1 km from the ADCP. This separation, particularly of the $N$ and $S$ data, raises concerns about the estimates of Ri that are addressed at the end of Section 3.3 and in the Appendix. The velocity, density and $\varepsilon$ data are averaged over the 12--30 min of the FLY bursts to provide the data from which $S$, $N$ and $\varepsilon$ are derived, characterising the flow every hour. Figure 1 shows the means of all these hourly values in each sea, omitting the regions near sea surface and seabed that are excluded from analysis. The notation $<$ $>$ implies time averages for each sea. Averages for the full sets of data (solid lines) and those including only data sets in which disturbances are unstable (dashed lines) are shown separately. Generally, $<$Ri$>$ = $<$$N^2$$>$$/$$<$$S^2$$>$ is reduced when averaged only over periods in which disturbances are unstable, but $<$$\varepsilon$$>$, the mean value of the rate of dissipation of turbulent kinetic energy per unit mass, is hardly altered.

L10 examines 24 hourly values of averaged flows in the Clyde Sea. The site (Table 1) is a few kilometres from the broad crest of the 40 m deep sill at the entrance of the fjord, the source of M$_2$ internal tides that dominate the shear at mid-depth in the water column. Bursts of 6 FLY casts made within 12--18 min every hour provide
profiles of density and $\varepsilon$ with 1 m vertical resolution. Mean density and $\varepsilon$ profiles are obtained by averaging the data from the hourly bursts. Mean velocity profiles, averaged over the same time periods (i.e., that of the bursts), are obtained from ADCP data sampled at 2 m vertical intervals. This provides one mean velocity and density profile every hour from which $\textrm{Ri}(z)$ is calculated. In two of the hourly periods, $\textrm{Ri}_{\textrm{min}}>\sfrac{1}{4}$, implying that the flow is stable. In the other periods, the growth rates are estimated as described by L10. That is, the growth rates are estimated either directly by solving the T--G equation or, where the solutions do not converge as the vertical scale of the grid on which velocities and buoyancy frequencies are estimated decreases, by extrapolation from solutions in which the mean flow has been increased by a factor $(1+\Phi)$.

Figure 1a shows averaged data from the Clyde Sea. The mean shear, $<$$|S|$$>$, has a value near $1.8\times10^{-2}$ $ \textrm{s}^{-1}$. The mean buoyancy frequency, $<$$N$$>$, is typically about $1.5\times10^{-2}$ $\textrm{s}^{-1}$ with no narrow maximum; the mean data show no pronounced pycnocline. The inverse $<$Ri$>$ is about 2; mean values of $<$Ri$>$ = $<$$N^2$$>$$/$$<$$|S|^2$$>$ $\approx$ $\sfrac{1}{2}$. Dissipation rate are of order $10^{-7}$ W kg$^{-1}$, substantially larger than in the other seas (Figs. 1b and 1c).

Observations in the summer stratified region of the western Irish Sea were made in 2006 as described by \citet{Green10}. The flow is dominated by the barotropic and baroclinic modes 1--3 semi-diurnal tides, with the baroclinic modes apparently being generated where the bottom slope is critical to the M$_2$ frequency west of the Isle of Man, some 60 km from the observation site. Inertial motions are relatively weak. Data are obtained from a moored thermistor chain, a 300 kHz ADCP (2 m vertical bin size), CTD casts and a 50 hr series of FLY turbulence yo-yo casts, in hourly bursts taking about 25--30 min. Mean profiles are shown in Fig.1b. There is a distinct peak in the shear at a height of about 78 m above the seabed that results in smaller values of $<$Ri$>$ but no
evident corresponding peak in dissipation, $<$$\varepsilon$$>$, which is about an order of magnitude less than in the Clyde Sea. Data obtained only during the periods in which small disturbances to hourly averaged samples are unstable are shown by dashed lines. This selected data have greater peak values of $<$$|S|$$>$ and $<$Ri$>$$^{-1}$ at 78 m with $<$Ri$>$ about $\sfrac{1}{3}$, although $<$$\varepsilon$$>$ falls slightly in size at this level.

\citet{Palmer08} describe the data set (referred to as CS3.2) obtained in the Celtic Sea in 2003 remote from regions where topographic generation of internal tides may be significant. They used a 300 kHz ADCP with 2 m vertical resolution, moored thermistor chains, CTD and FLY turbulence yo-yo series, hourly bursts again taking 25--30 min. The shear in the pycnocline is dominated by contributions from a westward-propagating diurnal baroclinic tide, possibly generated in the Bristol Channel some 150 km away to the east, and by near-inertial oscillations remaining from a wind burst 5 days prior to the observations. The mean data are shown in Fig.1c. There is a pycnocline with enhanced shear at about 82 m above the seabed. Dissipation is similar to that in the Irish Sea, and $<$Ri$>$ generally exceeds 1, but is reduced to $\sfrac{1}{2}$ during unstable periods (dashe lines in Fig.1c).

\section{Analysis and results}
Analysis of the hourly data of the Irish and Celtic Seas is conducted as in L10. The growth rate of the fastest growing disturbances is denoted as $kc_i$ and the buoyancy period as $T_b$, equal to $2\pi/N$, where $N$ is the buoyancy frequency of the hourly mean flow at the level of maximum displacement of the fastest growing wave mode. All values of Ri are based on the interpolated 0.2 m data. Results for the two seas are summarized in Tables 2 and 3. These may be compared with those for the Clyde Sea given in a similar table by L10 (his Table 1].

\subsection{Statistics of the Richardson number}

Figure 2 is a histogram of Ri$_{\textrm{min}}$ based on the interpolated data with a vertical resolution of 0.2 m, and Table 4 summarizes values of Ri$_{\textrm{min}}$, etc., in the three seas. The majority of very small values are found in the Clyde Sea (Fig. 2a), with none of the relatively large values exceeding 0.30 found in the Celtic and Irish Seas. Only a fraction, 8\%, of the Clyde Sea Ri$_{\textrm{min}}$ values lie in the range 0.21 to 0.30 surrounding $\sfrac{1}{4}$, whilst the Celtic and Irish Seas have corresponding fractions of 22\% and 33\%, respectively. There is however no clear evidence of a maximum near $\sfrac{1}{4}$. Twenty of the 96 hourly periods for the three seas are stable even though Ri$_{\textrm{min}}$ is less than $\sfrac{1}{4}$; about 37\% of the 54 flows in which Ri$_{\textrm{min}} $$<$ $\sfrac{1}{4}$ are found to be stable. The mean value of Ri$_{\textrm{min}}$ in the Clyde Sea is less than 0.1 and smaller than the mean values in the Irish and Celtic Seas, each of which exceeds $\sfrac{1}{4}$.

We define a `marginal range' of stability as that in which $0.8<\textrm{Ri}_{\textrm{min}}/\textrm{Ri}_c<1.2$, where the observed mean flow is close to critical (where $kc_i = 0$). This definition is consistent with the concept of `marginal instability' defined by \cite{Thorpe09b}, and compares with a similar concept proposed by \cite{Smyth13a} in their study of deep cycle turbulence in the eastern equatorial Pacific ocean. Figure 3 is the histogram of Ri$_{\textrm{c}}$ for all the flows in which $\textrm{Ri}_{\textrm{min}}/\textrm{Ri}_{\textrm{c}}<1.2$, i.e., including the marginal range, but with the unstable flows ($\textrm{Ri}_{\textrm{min}}<\textrm{Ri}_{\textrm{c}}$) being separated from the stable flows
(Ri$_{\textrm{min}}\ge$ Ri$_{\textrm{c}}$). Sixty percent of values of Ri$_{\textrm{c}}$ lie between 0.21 and 0.25, but with a substantial `tail' extending to the smallest resolved values. Forty-two percent of the unstable hourly flows have Ri$_{\textrm{c}}$ $<$ 0.21. It is evident from Fig. 3 that, as found earlier [e.g., by {\textit{Galperin et al.}}, 2007, and L10], the selection of $\sfrac{1}{4}$ as a critical gradient Richardson number is not generally valid. 

Figure 4 shows the distribution of $\textrm{Ri}_{\textrm{min}}/\textrm{Ri}_c$ ($<$ 1.2) in each of the three seas. Overall, a proportion of 36\% of the flows are unstable and 31\% are in the marginal range. As shown in Table 4, the Clyde Sea has a greater proportion (63\%) of unstable hourly periods (with $\textrm{Ri}_{\textrm{min}}/\textrm{Ri}_c$ $<$ 1) than the Irish Sea (30\%) and the Celtic Sea (17\%). The Clyde Sea has also the greatest proportion of flows that are in the marginal range (42\%), and relatively fewer that are stable with $\textrm{Ri}_{\textrm{min}}/\textrm{Ri}_c$ $>$ 1.2.

\subsection{Variability of the exponential growth period}
The exponential growth period of small unstable disturbances, $\tau$, is equal to the inverse of the growth rate, $kc_i$. We non-dimensionalize $\tau$ with the buoyancy period $T_b$. Figure 5 shows the variation of log$_{\textrm{10}}(\tau/T_b)$ with $\textrm{Ri}_{\textrm{min}}/\textrm{Ri}_{\textrm{c}}$. The scatter is roughly equal to the uncertainty indicated by the error bars, but there is consistency of the values determined in the three seas, and a fairly tight relationship between the two variables. In general, the non-dimensional growth rate increases as Ri$_{\textrm{min}}$ ($<$ Ri$_{\textrm{c}}$) decreases. Although approximately linear over much of the $\textrm{Ri}_{\textrm{min}}/\textrm{Ri}_{\textrm{c}}$ range, there is a rise of values of log$_{\textrm{10}}(\tau/T_b)$ beyond the linear trend as $\textrm{Ri}_{\textrm{min}}/\textrm{Ri}_{\textrm{c}}$ tends to unity. This is expected: the growth time, $\tau$, must tend to infinity as Ri$_{\textrm{min}}$ tends towards stable conditions at Ri$_{\textrm{min}}$ $=$ Ri$_{\textrm{c}}$. It is worthwhile noting that the relationship between log$_{\textrm{10}}(\tau/T_b)$ and $\textrm{Ri}_{\textrm{min}}/\textrm{Ri}_{\textrm{c}}$ is in fact quite similar to that for a simple (hyperbolic tangent) shear layer (Fig. 5; see also \cite{Hazel72}, his Figure 1). 

The averaging times of 12--30 min can now be compared to $T_b$ and $\tau$. Since $N$ is typically about $1.5\times10^{-2}$ s$^{-1}$ (Fig. 1), the buoyancy period is about 7 min. Internal waves have periods that exceed the buoyancy period. Only fluctuations caused by relatively high-frequency waves and the (higher frequency) turbulent motions are averaged; variations caused by the baroclinic M$_2$ tides or inertial waves that dominate the flow variations are not averaged, but resolved. The largest values of $kc_i$ (Tables 2 and 3, and L10) are about 10$^{-2}$ s$^{-1}$ (the smallest being near-zero; stable conditions), and so $\tau$ is greater than about 1 min. Those disturbances with $\tau$ near this limit may grow and become unstable, perhaps producing turbulence, during the sampling period. When $\tau$ is roughly equal to the averaging period (i.e., when $kc_i = 5\times10^{-4}$ s$^{-1}$ $-1.4\times10^{-3}$ s$^{-1}$), unstable disturbances will grow rather little during the averaging periods and the flow will be little modified by their presence. An alternative view is that if the collapse time of turbulence, $T_{coll}$, is less than $T_b$, turbulence must be generated during the sampling periods if it is to be sustained. The value of $T_{coll}$ is uncertain, in part because it depends on the mean Ri and on what characteristic of turbulence is selected. \citet{Smyth97} suggest \textit{e}-folding times of ($4.3\pm1.8$)$N^{-1}$, whilst, in numerical experiments in the absence of a mean shear,
\citet{Staquet98} find collapse of the vertical heat flux and the onset of anisotropy in a time of about 9.4$N^{-1}$. In a uniformly stratified turbulent flow with uniform mean shear, \citet{Diamessis04} examine the times for the volume fraction of statically unstable regions (`overturns') to decay to zero. The times depend on the mean flow gradient Richardson number, $<$Ri$>$, being about $4N^{-1}$ in the absence of shear, and about $6N^{-1}$ when $<$Ri$>$ = 0.5, with an even greater decay time when $<$Ri$>$ = 0.2. (There appears, however, to be no collapse if $<$Ri$>$ = 0.05.) If we accept Smyth et al.'s estimate, the time $T_{coll} \sim 3.5-7$ min, and, to be sustained, turbulence must generally be generated during the averaging periods.

\subsection{Correlation of turbulent dissipation with instability}
We turn next to the turbulent dissipation. The mean dissipation near the level at which the amplitude of small unstable disturbances is greatest is quantified as $<$$\varepsilon$$>$ = $E/(z_2-z_1)$, where $E = \int_{z_1}^{z_2} \! \varepsilon \, dz$ and $\varepsilon$ is the rate of loss of turbulent kinetic energy per unit mass and $z_1$ to $z_2$ is the vertical interval over which the modulus of the amplitude of the disturbances exceeds 20\% of its maximum value. Taking $N$ as the averaged buoyancy frequency in the range $z_1$ to $z_2$, and a nominal value, 10$^{-6}$ m$^2$ s$^{-1}$, for the molecular viscosity, $\nu$, Figs. 6--8 show the variation of log$_{\textrm{10}}($$<$$\varepsilon$$>$$/\nu N^2)$ (when unstable disturbances are found) with three parameters that quantify instability: Ri$_{\textrm{min}}$ (for the unstable cases), $\textrm{Ri}_{\textrm{min}}/\textrm{Ri}_{\textrm{c}}$ and log$_{\textrm{10}}(\tau/T_b)$, respectively. Different symbols are used to denote data from the three seas. Although there is considerable scatter, there are evident trends in the normalized dissipation rates with each of the three other variables. The trends indicate that the greater is the measure of instability the higher is the dissipation. Equations for the linear regressions shown as dashed lines are shown in the figure captions. Given the correlation of $\textrm{Ri}_{\textrm{min}}/\textrm{Ri}_{\textrm{c}}$ and log$_{\textrm{10}}(\tau/T_b)$ evident in Fig.5, the similarity of trends in Figs.7 and 8 is as expected. The three figures support the hypothesis that turbulent dissipation is partly related to the instability of the flow.

The regression line in Fig.7 indicates that $<$$\varepsilon$$>$$/\nu N^2\approx85$ to within a multiplicative factor of about 3, when $\textrm{Ri}_{\textrm{min}}/\textrm{Ri}_{\textrm{c}}$  = 1 (i.e., when there is zero growth rate of small disturbances), suggesting that, during the periods of instability, there is a ‘background’ level of turbulence of about $<$$\varepsilon$$>$ $\approx85\nu N^2 = 1.91\times10^{-8}$ W kg$^{-1}$, taking $N$ as its mean value
of about $1.5\times10^{-2}$ s$^{-1}$. This compares with a background level of dissipation of $(6.72\pm0.22)\times10^{-8}$ W kg$^{-1}$ found by \citet{Liu10} for the Clyde Sea. As for the Irish Sea, although
given the large scatter of data points it is hard to obtain a reliable estimate of the background dissipation rate, it is evident from Fig.7 that it must be less than $1.91\times10^{-8}$ W kg$^{-1}$ because most of the data points are below the regression line. The 4 points from the Celtic Sea are too few to give a clear indication of the background dissipation rate there. The data imply, however, that there is generally a background level of turbulent dissipation that is augmented by flow instability; instability of the mean flow does not account for all the dissipation.

Figure 8 appears to imply a smaller power law relation between $<$$\varepsilon$$>$ and the growth
rate (i.e., $<$$\varepsilon$$>$ $\propto(kc_i)^{0.57}$) for the three seas taken together than is found by L10 (i.e., $<$$\varepsilon$$>$ $\propto(kc_i)^{1.83}$) for the Clyde Sea alone. The parameter $I$ = $<$$\varepsilon$$>$$/\nu N^2$ can be regarded as an index of the isotropy of the turbulent motion at the level of maximum disturbance amplitude; \citet{Gargett84} find that turbulence is isotropic when $I$ $>$ 200, whilst if $I$ $<$ $O$(20)
turbulence produces no significant buoyancy flux \citep{Stillinger83,Itsweire93}. The level of the mean local dissipation rate is generally sufficient to promote a significant buoyancy flux, but not always to support isotropic turbulence. During the periods of observation, turbulence in the Clyde Sea at the location of maximum disturbance amplitude is generally isotropic whilst that in the Irish Sea is not.

The separation between the sites of the ADCPs and the FLY measurements of $N$ and $\varepsilon$, mentioned in Section 2, is of concern, for it implies that the measured values of the terms $N^2$ and $S^2$ that appear in Ri are not horizontally collocated. (Measurements of $\varepsilon$ and $N$ are, however, both obtained by the FLY and collocated, so that $I$ is not affected by separation.) Implicit in our calculation of Ri is that the temporal mean values of $S$ taken over the duration of the FLY bursts are well correlated over the distance, $D$, between the ADCPs and the FLY sites, but no estimates of the horizontal scales over which values of $S$ are correlated are available to prove this. Whilst the mean flows of order 0.1 m s$^{-1}$ imply (making a Taylor hypothesis) that the temporal averages over 12--30 min may represent horizontal averages of about 70--200 m, this is hardly sufficient to span the distance, $D$, or to convincingly demonstrate that the values of Ri at the FLY site are accurately estimated using the relatively distant ADCP data. The main cause of variation in profiles is, however, the internal M$_2$ tide which has phase speeds of typically 0.5 m s$^{-1}$, so transmitting variations over the $O$(1 km) separation between sampling sites in about 40 min, greater than the averaging times, and not short enough for the temporal average to remove doubt about the site separation. Contributing effects to a vertical offset in the $S$ and $N$ measurements in the Clyde Sea and the uncertainty in estimates of Ri are discussed in the Appendix.

\subsection{Marginal instability and the wave--turbulence boundary}
We return now to the topic of marginal stability referred to in Section 1. The idea that the ocean and some other naturally occurring dynamical systems are in a state of marginal instability is not new. It dates back to speculations by \citet{Malkus56} (see also Lumley 1981) and may be related to the concept of `self-organized criticality' \citep{Bak88}. It has, for some 40 years, been supposed that the internal wave fields of the ocean and atmosphere are in a marginal `saturated' state \citep [e.g.,][]{Staquet02}, any additional supply of energy beyond that of the Garrett--Munk universal spectrum being quickly redistributed in frequency and wavenumber through resonant interactions, and cascaded to those small vertical scales where it can be dissipated in breaking events. This formulation of `marginality' in a mean flow that has a large gradient Richardson number is \textit{non-linear} in that it involves a cascade of energy through triad or higher order wave interactions to scales at which energy is removed from the system by wave breaking.

The alternative, \textit{strictly linear}, view of marginality adopted here depends on the stability of observed mean stratified shear flows to small disturbances, a formulation of marginal stability that goes back to the theorem of \citet{Miles61} and \citet{Howard61} that steady, inviscid, non-diffusive, horizontally moving, stratified shear flows are stable to small disturbances if the gradient Richardson number, $\textrm{Ri}(z)$, is greater than or equal to $\sfrac{1}{4}$ everywhere in the fluid. The condition Ri$_{\textrm{min}}$ $<$ $\sfrac{1}{4}$ has often been taken as being a \textit{sufficient} condition for instability of a flow, whether steady or not \citep [e.g.,][]{Polzin96}, and even that the flow will become turbulent if Ri$_{\textrm{min}}$ $<$ $\sfrac{1}{4}$ \citep [e.g.,][]{Rippeth05b}. It is
however already evident from Fig.3 and earlier studies by \citet{Thorpe07,Thorpe09a}, \citet{Thorpe09b} and L10, that Ri$_{\textrm{c}}$ is often less than $\sfrac{1}{4}$. Marginal conditions in which Ri$_{\textrm{min}}$ is close to Ri$_{\textrm{c}}$ are not always common (Table 4). The proportion of time in which flows are marginal or unstable differs in data sets from the three seas, with no evidence of a simple law or parameterization governing the distribution of $\textrm{Ri}_{\textrm{min}}/\textrm{Ri}_{\textrm{c}}$ or unifying the state of the flow. A `background' level of turbulence is apparent. Evidently some measure of internal wave activity may be required to formulate or parameterize the state of flow or its mixing. 

The distinction between the two formulations of marginality is addressed by \citet{DAsaro00} (see also Baumert and Peters 2009) in a discussion of what they term `The Wave--Turbulence (W--T) Transition'. Internal waves can be separated from turbulence because their frequency measured in a Lagrangian frame of reference is less than the buoyancy frequency $N$. \citet{DAsaro00} model the spectrum and point to the different physics in the two regimes -- each of which may be in a state of marginal stability, but with different characteristics: the internal wave regime is below the W--T boundary, characterised by large overall Ri with patchy mixing depending on an energy cascade controlled by wave--wave and possibly vortical mode interactions, and the relatively turbulent regime above the W--T boundary at lower overall Ri where mixing is controlled by instabilities of the mean flow. The W--T transition marks a change from
energy transfer controlled by wave--wave interaction from a large to a small dissipation scale to one controlled by the instability of the mean flow (as examined above in the three seas) and consequent turbulence. \citet{DAsaro00} propose an approximate relation to determine the least rate of dissipation of turbulent kinetic energy required for
conditions to be above the W--T transition:
 \begin{equation}
 \varepsilon_{trans}=f(NH/2\pi)^2/2
 \end{equation}
where $f$ is the Coriolis parameter and $H$ is the water depth. In the deep ocean, dissipation rates are relatively low and mixing is dominated by occasional wave breaking with mean
conditions and dissipation rates generally well below the W--T transition predicted by (1). Flows in the relatively shallow water of the continental shelf have higher dissipation rates, possibly with $<$$\varepsilon$$>$ $>$ $\varepsilon_{trans}$. Parameterizations that ignore the effects of internal waves are appropriate only above the W$-$T transition. A thorough discussion and comparison of the various parametrizations is given by \citet{MacKinnon05}.

Where, in relation to the W--T boundary, do the data from the three seas lie? Figure 9 shows the variation of log$_{10}($$<$$\varepsilon$$>$$/\varepsilon_{trans})$ with $\textrm{Ri}_{\textrm{min}}/\textrm{Ri}_{\textrm{c}}$, where $\varepsilon_{trans}$ is given by (1) and values of $f$ and $H$ are given in Table 1. With the exception of one point, $<$$\varepsilon$$>$ is less
than $\varepsilon_{trans}$, implying data lie below the W--T transition. For several reasons, however, the ratio, $<$$\varepsilon$$>$$/\varepsilon_{trans}$, is at best only an approximate guide. As explained by \citet{DAsaro00}, (1) is an approximate estimate of the dissipation rate of turbulent kinetic energy at the W--T boundary, and it depends on assumptions related to the rate of decay of turbulent kinetic energy and the form of the vertical wavenumber spectrum near the W--T boundary. The term $NH/2\pi$ represents the speed of the lowest internal wave mode (as selected by \citet{DAsaro00}: it is actually that of the second mode in a layer of depth $H$ with uniform stratification, $N$, with no mean shear) and will differ in the non-uniform mean density gradient and non-zero mean shear of the three seas. Our selection of data in Fig.9 is confined to those with unstable modes, the periods expected to have greater $<$$\varepsilon$$>$ and therefore to exceed the W$-$T boundary. The values of $N$ and $<$$\varepsilon$$>$ are those surrounding the maximum disturbances.

A difference between the Clyde and Irish Sea data is evident however in Fig.9; the Clyde Sea has values of $<$$\varepsilon$$>$ $>$ $\varepsilon_{trans}$ that increase as $\textrm{Ri}_{\textrm{min}}/\textrm{Ri}_{\textrm{c}}$ decreases (i.e., as unstable conditions become more prevalent), and that are almost an order of magnitude greater than those in the Irish Sea (and on average greater than the sparse data from the Celtic Sea). The Clyde Sea appears therefore to be near the W--T boundary, whilst the Irish Sea is below it. This is consistent with the evidence from the histograms of Ri$_{\textrm{min}}$ and $\textrm{Ri}_{\textrm{min}}/\textrm{Ri}_{\textrm{c}}$, and the variation of $<$$\varepsilon$$>$$/\nu N^2$ with $\textrm{Ri}_{\textrm{min}}/\textrm{Ri}_{\textrm{c}}$ and $\tau/T_b$. It may be expected that dissipation rates will only be significantly related to the parameters that determine the stability of the mean flow (as in Fig.7 for the Clyde Sea) in conditions that exceed or are near to the W--T boundary. Well below it unquantified wave interactions lead to patchy turbulence that is less strongly related to the mean flow stability parameters (as for the Celtic and Irish Seas in Fig.7).

\section{Summary}
In this paper, the stability of flows observed at three different locations on the UK Continental Shelf are analyzed. The data, limited to particular times and isolated locations in each sea, are insufficient to draw general conclusions about how mixing varies between the three seas. However, using the analyzed hourly data (averaged over 12--30 min, about 2--4 times the buoyancy period):

\begin{enumerate} [(1)]

     \item further examples are found that show that Ri$_{\textrm{min}}$ $<$ $\sfrac{1}{4}$ is not a sufficient condition for instability: 37\% of the flows with Ri$_{\textrm{min}}$ $<$ $\sfrac{1}{4}$ are stable; Ri$_{\textrm{c}}$ $<$ 0.24 in 68\% of the cases where Ri$_{\textrm{min}}$ $<$ $\sfrac{1}{4}$;
     
      \item marginal conditions with 0.8 $\textless$ Ri$_{\textrm{min}}$$/$Ri$_\textrm{c}$ $\textless$ 1.2 occur in 28\% of the cases;
      
      \item values of $<$$\varepsilon$$>$$/\nu N^2$, a measure of the isotropy of the turbulence, show a decreasing trend with the stability parameters, Ri$_{\textrm{min}}$, Ri$_{\textrm{min}}$$/$Ri$_\textrm{c}$ and $\tau/T_b$ (Figs. 6--8); more turbulent mixing occurs as instability becomes more intense;
      
      \item in the periods of data analyzed, that near the sill in the Clyde Sea is most unstable, having a larger mean ratio, Ri$_{\textrm{min}}$$/$Ri$_\textrm{c}$, and greater values of the parameter, $<$$\varepsilon$$>$$/\varepsilon_{trans}$, relating the flow to the location of the W$-$T transition (possibly nearer the W$-$T transition than in the other seas);
      
      \item the data imply that there is a background level of dissipation that is augmented by instability; instability of the averaged flow does not account for all the turbulence observed in mid-water.

\end{enumerate}

The data do not support the hypothesis that the turbulent flows observed on the UK Continental Shelf adjust rapidly to conditions that are close to being marginal, or that flows in a particular location and period of time in one sea have stability characteristics that are very similar to those in another. The different statistical variations of Richardson
numbers in Figs. 2--4 indicate that no overall `law of mean flow marginality' covering the three seas is obeyed. 

The conclusions (4) and (5) above suggest that processes, possibly vortical modes or internal waves of relatively small period, provide a background level of turbulence augmented by flow instability. This points to one major
reservation in using the Taylor--Goldstein equation: the effects of turbulent viscosity and diffusivity are disregarded in examining the growth of small disturbances. These effects have been investigated in a more recent study \citep{Liu12}, and for the flows analyzed in this paper, they are found to be relatively small. However, in other cases, for example for the flows observed in the upper-equatorial Pacific, these effects seem to be crucial in the cycling of instability and the maintenance of relatively strong turbulence \citep{Smyth13}.

The present data are barely adequate, in particular because of the separation of the ADCP and FLY sites, and some of our conclusions (e.g., those based on Figs. 6--9) are subject to reservations. In future it is essential to collocate the density, $\varepsilon$ and velocity measurements.

%
\acknowledgments
Professor Tom Rippeth kindly provided available data from the three seas for analysis, and gave advice and permission for its use, for which I am most grateful. The data collection was partly supported by the British Natural Environmental Research Council through grant NE/F002432. This work was supported by the National Natural Science Foundation of China (41476006 and 41006017) and the Natural Science Foundation of Fujian Province of China (2015J06010). Professor Steve Thorpe is sincerely thanked for his substantial contributions to the work, and Professor Bill Smyth kindly drew my attention to the concept of `self-organized criticality'. 

%






%
%
%
\appendix
\section{The separation of FLY and ADCP data in the Clyde Sea}
We noted a possible mismatch in the locations of
high $S$ and high $N$ in the Clyde Sea data shown in Fig.1a: peaks in $<$$S$$>$ at heights above
bottom of 40 m and 12 m (figure column 1) are some 2 m lower than those in $<$$N$$>$
(column 2). (2 m is also the bin size of the ADCP measurements of current.) We are wondering whether it is because of this offset, the large $<$Ri$>$$^{-1}$ (column 3) are not related to any substantial increase in $<$$\varepsilon$$>$ (column 4).

No errors have been found in calculating and ascribing heights above bottom to the data. The raw ADCP data, in particular, were reworked to check the mean heights of the 2 m high data bins. Correction is made for the height of the ADCP above the bottom. The FLY data are referenced to the pressure at the seabed signalled by recorded tilt and pressure when the probe reaches the seabed.

Several effects might, however, contribute to a 2 m offset in $S$ and $N$ over the mean distance of 0.8 km between the ADCP and FLY locations in the Clyde Sea, the shift implying a slope of the constant $S$ and $N$ surfaces of 2 m in 0.8 km, or 0.0025:

\begin{enumerate} [(i)]

   \item \textbf{\textit{slope of the seabed}}. It is possible that some slight error in the ADCP heights could be introduced if the instrument were resting on a local mound or depression in the seabed, but this is unknown and is unlikely to amount to a change of more than 0.5 m. There is, however, a mean bottom slope from the sill to a shallow basin to the east of the Isle of Arran in the Clyde Sea estimated to be about 0.0027;
   
   \item \textbf{\textit{a geostrophic tilt of isopycnal surfaces.}} The horizontal density gradient, $\partial\rho/\partial x$, is calculated from the geostrophic or `thermal wind' equations using the vertical gradients of the tidally averaged velocity components. Dividing these by the vertical density gradients, $\partial\rho/\partial z$, the mean isopycnal slopes are found to be about 0.0025. (This does not however take into account any vertical offset of the data from the ADCP location, where $\partial\rho/\partial x$ is estimated, and the FLY location of $\partial\rho/\partial z$.)
   
   \item \textbf{\textit{the slope of isopycnals caused by the M$_2$ internal tide.}} The amplitude, $a$, of the internal tide at the site is about 8 m and its speed is about 0.5 m s$^{-1}$, giving a wavelength, $\lambda$, of about 22 km and a maximum isopycnal slope, $2\pi a/\lambda\sim 0.0022$,
or a root mean square (rms) slope (horizontally and vertically averaged, assuming a sinusoidal mode 1 variation) of about 0.0011.

\end{enumerate} 

Although each of these might contribute substantially to the slope associated with a perceived data shift of 2 m, the FLY casts were made around the ADCP location, both towards and away from the sill, and the three effects are unlikely to contribute much to the averaged data shown in Fig.1a.

As a test of the sensitivity of the estimates of Ri$_{\textrm{min}}$ to the possible discrepancies in the heights of the $S$ and $N$ data, offsets, $\Delta h$, ranging from $-2$ m to +2 m were imposed and the histogram of Fig. 2a was recreated. The results are shown in Fig.10. Although the distributions vary in detail:

\begin{enumerate}[a)]

   \item the number of hourly profiles with Ri$_{\textrm{min}}$ $>$ 0.25 (that must, therefore, be stable) is hardly affected; only at $\Delta h = +1$ m is the number increased from 2 to 3;

   \item the mean values of Ri$_{\textrm{min}}$ are 0.0856, 0.108, 0.097, 0.140 and 0.127 for $\Delta h=-2$ m, --1 m, 0 m, +1 m and +2 m, respectively. (The median values vary in a similar way and by similar amounts.) The changes in the mean Ri$_{\textrm{min}}$ are relatively small. More significantly, the rms differences of the hourly values of Ri$_{\textrm{min}}$ at $\Delta h=-2$ m, --1 m, +1 m and +2 m from those at zero offset are 0.032, 0.033, 0.033 and 0.035, respectively, implying an uncertainty of about 0.03, greater than a value, 0.005, estimated from the uncertainties in the measured $S$ and $N$; 

   \item in comparison with Figs. 2b and 2c, there is a much greater proportion of values of Ri$_{\textrm{min}}$ $<$ 0.25. In this sense and as concluded before, at the time and location of measurements, the flow in the Clyde Sea appears more likely to be more unstable than in the other seas. The sensitivity of log$_{10}($$<$$\varepsilon$$>$$/\nu N^2$) (see ordinates of Figs. 6--8) to the possible discrepancies in the heights of the $N$ and $\varepsilon$ FLY data from those of $S$ are assessed by finding revised values of $E$ through integrating $\varepsilon$ over ranges $z_1$ to $z_2$ offset by --2 m, --1 m, +1 m and +2 m, and of $N$ by using the same offset values. The rms variations of these hourly differences from those at zero offset are 0.140, 0.072, 0.085 and 0.146, respectively. These are smaller than the uncertainty of 0.6 shown in the figures. 
 
\end{enumerate}

We conclude that the variations of scaled values of $<$$\varepsilon$$>$ with measures of the flow stability shown in Figs. 7 \& 8 are statistically \textit{significant} even though the estimates of Ri may be wanting in their precision.
\bibliographystyle{ametsoc2014}
\bibliography{references}

\begin{thebibliography}{27}
\providecommand{\natexlab}[1]{#1}
\providecommand{\url}[1]{\texttt{#1}}
\renewcommand{\UrlFont}{\rmfamily}
\providecommand{\urlprefix}{URL }
\expandafter\ifx\csname urlstyle\endcsname\relax
  \providecommand{\doi}[1]{doi:\discretionary{}{}{}#1}\else
  \providecommand{\doi}{doi:\discretionary{}{}{}\begingroup
  \urlstyle{rm}\Url}\fi
\providecommand{\eprint}[2][]{\url{#2}}

\bibitem[{Bak et~al.(1988)Bak, Tang,, and Wiesenfeld}]{Bak88}
Bak, P., C.~Tang, and K.~Wiesenfeld, 1988: Self-organised criticality.
  \textit{Phys. Rev. A}, \textbf{38}, 364--374.

\bibitem[{D'Asaro and Lien(2000)D'Asaro, and Lien}]{DAsaro00}
D'Asaro, E.~A., and R.~C. Lien, 2000: The wave--turbulence transition for
  stratified flows. \textit{J. Phys. Oceanogr.}, \textbf{30}, 1669--1678.

\bibitem[{Diamessis and Nomura(2004)Diamessis, and Nomura}]{Diamessis04}
Diamessis, P.~J., and K.~K. Nomura, 2004: The structure and dynamics of
  overturns in stably stratified homogeneous turbulence. \textit{J. Fluid
  Mech.}, \textbf{499}, 197--229.

\bibitem[{Gargett et~al.(1984)Gargett, Osborn,, and Nasmyth}]{Gargett84}
Gargett, A.~E., T.~R. Osborn, and P.~W. Nasmyth, 1984: Local isotropy and the
  decay of turbulence in a stratified fluid. \textit{J. Fluid Mech.},
  \textbf{144}, 231--280.

\bibitem[{Green et~al.(2010)Green, Simpson, Thorpe,, and Rippeth}]{Green10}
Green, J. A.~M., J.~H. Simpson, S.~A. Thorpe, and T.~P. Rippeth, 2010:
  Observations of internal waves in the isolated seasonally stratified region
  of the western irish sea. \textit{Cont. Shelf Res.}, \textbf{30}, 214--225.

\bibitem[{Hazel(1972)}]{Hazel72}
Hazel, P., 1972: Numerical studies of the stability of inviscid stratified
  shear flows. \textit{J. Fluid Mech.}, \textbf{51}, 39--61.

\bibitem[{Howard(1961)}]{Howard61}
Howard, L.~N., 1961: Note on a paper by John W. Miles. \textit{J. Fluid Mech.},
  \textbf{10}, 509--512.

\bibitem[{Itsweire et~al.(1993)Itsweire, Koseff, Briggs,, and
  Ferziger}]{Itsweire93}
Itsweire, E.~L., J.~R. Koseff, D.~A. Briggs, and J.~H. Ferziger, 1993:
  Turbulence in stratified shear flows: implications for interpreting
  shear-induced mixing in the ocean. \textit{J. Phys. Oceanogr.}, \textbf{23},
  1508--1522.

\bibitem[{Liu(2010)}]{Liu10}
Liu, Z., 2010: Instability of baroclinic tidal flow in a stratified fjord.
  \textit{J. Phys. Oceanogr.}, \textbf{40}, 139--154.

\bibitem[{Liu et~al.(2012)Liu, Thorpe,, and Smyth}]{Liu12}
Liu, Z., S.~A. Thorpe, and W.~D. Smyth, 2012: Instability and hydraulics of
  turbulent stratified shear flows. \textit{J. Fluid Mech.}, \textbf{695},
  235--256.

\bibitem[{MacKinnon and Gregg(2005)MacKinnon, and Gregg}]{MacKinnon05}
MacKinnon, J.~A., and M.~C. Gregg, 2005: Spring mixing: turbulence and internal
  waves during restratification on the new england shelf. \textit{J. Phys.
  Oceanogr.}, \textbf{35}, 2425--2443.

\bibitem[{Malkus(1956)}]{Malkus56}
Malkus, W. V.~R., 1956: Outline of a theory of turbulent shear flow. \textit{J.
  Fluid Mech.}, \textbf{1}, 521--539.

\bibitem[{Miles(1961)}]{Miles61}
Miles, J., 1961: On the stability of heterogeneous shear flows. \textit{J.
  Fluid Mech.}, \textbf{10}, 496--508.

\bibitem[{Monserrat and Thorpe(1996)Monserrat, and Thorpe}]{Monserrat96}
Monserrat, S., and A.~J. Thorpe, 1996: Use of ducting theory in an observed
  case of gravity waves. \textit{J. Atmos. Sci.}, \textbf{53}, 1724--1736.

\bibitem[{Palmer et~al.(2008)Palmer, Rippeth,, and Simpson}]{Palmer08}
Palmer, M.~R., T.~P. Rippeth, and J.~H. Simpson, 2008: An investigation of
  internal mixing in a seasonally stratified shelf sea. \textit{J. Geophys.
  Res.}, \textbf{113}, C12\,005.

\bibitem[{Polzin(1996)}]{Polzin96}
Polzin, K., 1996: Statistics of the richardson number: mixing models and
  finestructure. \textit{J. Phys. Oceanogr.}, \textbf{26}, 1409--1425.

\bibitem[{Rippeth(2005)}]{Rippeth05a}
Rippeth, T.~P., 2005: Mixing in seasonally stratified shelf seas: a shifting
  paradigm. \textit{Phil. Trans. R. Soc. A.}, \textbf{363}, 2837--2854.

\bibitem[{Rippeth et~al.(2005)Rippeth, Palmer, Simpson, Fisher,, and
  Sharples}]{Rippeth05b}
Rippeth, T.~P., M.~R. Palmer, J.~H. Simpson, N.~R. Fisher, and J.~Sharples,
  2005: Thermocline mixing in summer stratified continental shelf seas.
  \textit{Geophys. Res. Lett.}, \textbf{32}, L05\,602.

\bibitem[{Smyth and Moum(2013)Smyth, and Moum}]{Smyth13a}
Smyth, W.~D., and J.~N. Moum, 2013: Marginal instability and deep cycle
  turbulence in the eastern equatorial pacific ocean. \textit{Geophys. Res.
  Lett.}, \textbf{40}, 6181--6185.

\bibitem[{Smyth et~al.(2013)Smyth, Moum, Li,, and Thorpe}]{Smyth13}
Smyth, W.~D., J.~N. Moum, L.~Li, and S.~A. Thorpe, 2013: Diurnal shear
  instability, the descent of the surface shear layer, and the deep cycle of
  equatorial turbulence. \textit{J. Phys. Oceanogr.}, \textbf{43}, 2432--2455.

\bibitem[{Smyth et~al.(1997)Smyth, Zavialov,, and Moum}]{Smyth97}
Smyth, W.~D., P.~O. Zavialov, and J.~N. Moum, 1997: Decay of turbulence in the
  upper ocean following sudden isolation from surface forcing. \textit{J. Phys.
  Oceanogr.}, \textbf{27}, 810--822.

\bibitem[{Staquet and Godeferd(1998)Staquet, and Godeferd}]{Staquet98}
Staquet, C., and F.~S. Godeferd, 1998: Statistical modelling and direct
  numerical simulations of decaying stably stratified turbulence. part 1. flow
  energetics. \textit{J. Fluid Mech.}, \textbf{360}, 295--340.

\bibitem[{Staquet and Sommeria(2002)Staquet, and Sommeria}]{Staquet02}
Staquet, C., and J.~Sommeria, 2002: Internal gravity waves: from instabilities
  to turbulence. \textit{Annu. Rev. Fluid Mech.}, \textbf{34}, 559--593.

\bibitem[{Stillinger et~al.(1983)Stillinger, Helland,, and van
  Atta}]{Stillinger83}
Stillinger, D.~C., K.~N. Helland, and C.~W. van Atta, 1983: Experiments on the
  transition of homogeneous turbulence to internal waves in a stratified fluid.
  \textit{J. Fluid Mech.}, \textbf{131}, 91--122.

\bibitem[{Thorpe and Liu(2009)Thorpe, and Liu}]{Thorpe09b}
Thorpe, S.~A., and Z.~Liu, 2009: Marginal instability? \textit{J. Phys.
  Oceanogr.}, \textbf{39}, 2373--2381.

\bibitem[{Thorpe and Ozen(2007)Thorpe, and Ozen}]{Thorpe07}
Thorpe, S.~A., and B.~Ozen, 2007: Are cascading flows stable? \textit{J. Fluid
  Mech.}, \textbf{589}, 411--432.

\bibitem[{Thorpe and Ozen(2009)Thorpe, and Ozen}]{Thorpe09a}
Thorpe, S.~A., and B.~Ozen, 2009: Corrigendum: Are cascading flows stable?
  \textit{J. Fluid Mech.}, \textbf{631}, 441--442.

\end{thebibliography}

%

\begin{table}
\renewcommand\arraystretch{1.5}
\caption{Values characterising the three seas: dates of observations, water depth,
latitude, Coriolis parameter, location in the spring-neap tidal cycle and the maximum
depth averaged tidal flow.}
\centering
\begin{tabular}{l c c c}
\hline
Parameter  & Clyde Sea  & Irish Sea  & Celtic Sea \\
\hline
  Dates  & July 1--2 2002 & July 16--18 2006 & August 10--11 2003  \\
  Depth, $H$ (m)  & 58 & 103 & 95 \\
  Latitude  & 55$^{\circ}$21$'$N & 53$^{\circ}$42$'$N & 51$^{\circ}$28$'$N \\
  Coriolis parameter, $f$ (s$^{-1}$)  & 1.196$\times$10$^{-4}$ & 1.172$\times$10$^{-4}$ & 1.138$\times$10$^{-4}$ \\
  Days after spring tide & 5 & 2 & 8\\
  Maximum mean tidal flow (m s$^{-1}$) & 0.21 & 0.47 & 0.37\\
\hline
\end{tabular}
\end{table}

\clearpage
\begin{table}
\renewcommand\arraystretch{1.5}
\caption{Summary of analysis from the Irish Sea. Here, $\lambda$ and $kc_i$ are respectively the
wavelength and growth rate of the fastest growing disturbance, and $N$ is the buoyancy
frequency at the level where the amplitude of the streamfunction is maximum; $\tau = (kc_i)^{-1}$,
is the $e$-folding period of the fastest growing disturbance, and $T_b = 2\pi N^{-1}$ is the buoyancy
period at the level where the amplitude of the streamfunction is maximum. The marginal
instability parameter $\Phi_{\textrm{c}}$ and the critical gradient Richardson number Ri$_{\textrm{c}}$ of the flows are
also listed. The uncertainties in $\Phi_{\textrm{c}}$ are estimated to be about 0.02, or less, and in Ri$_{\textrm{c}}$
about 0.03.}
\centering
\begin{tabular}{c c c c c c c c}
\hline
Hour  & Mode  & $\lambda$ (m)  & $kc_i$ (s$^{-1}$)  & $N$ (s$^{-1}$) & $\tau/T_b$ & $\Phi_c$ & Ri$_c$\\
\hline
0 & 1 & 29.5 & 0.97$\times$10$^{-3}$ & 1.04$\times$10$^{-2}$ & 1.71 & --0.13 & 0.20\\
1 & -- &  -- & 0 (stable)  & --  & --  & 0.10 & 0.20\\
2 & 1 & 30.0 & 0.45$\times$10$^{-3}$  & 1.40$\times$10$^{-2}$  & 4.94 & --0.05 & 0.25\\
3 & 1 & 24.5 & 2.68$\times$10$^{-3}$ & 1.15$\times$10$^{-2}$ & 0.69 & --0.26 & 0.25\\
4 & -- &  -- &  -- &  -- &  -- & 0.00 & 0.22\\
5 & 1 & 20.0 & 0.67$\times$10$^{-3}$ & 1.60$\times$10$^{-2}$ & 20.30 & --0.05 & 0.25\\
6--8$^*$ & -- & -- & -- & -- & -- & -- & --\\
9 & -- & -- & 0 (stable) & -- &  -- & 0.54 & 0.09\\
10 & -- &  -- & 0 (stable) & -- &  -- &  0.02 & 0.25\\
11 & 2 & 25.0 & 1.11$\times$10$^{-3}$ & 1.93$\times$10$^{-2}$ & 2.77 & --0.15 & 0.24\\
12 & -- &  -- & 0 (stable) & -- & -- & 0.12 & 0.23\\
13 & -- &  -- &  0 (stable) &  -- &  -- & 0.14 & 0.22\\
14 & 1 & 29.5 & 2.30$\times$10$^{-3}$ & 1.21$\times$10$^{-2}$ & 0.84 & --0.23 & 0.24\\
15 & -- &  --  & -- &  --  & -- & 0.09 & 0.22\\
16 & -- &  -- &  -- &  -- &  -- &  0.09 & 0.22\\
17 & 1  & 29.0 & 2.78$\times$10$^{-3}$ & 1.45$\times$10$^{-2}$ & 0.82 & --0.31 & 0.25\\
18--22$^*$ & -- & -- & -- & -- & -- & -- & --\\
23 & -- &  -- &  0 (stable) & -- &  -- & 0.42 & 0.10\\
24$^*$ & -- & -- & -- & -- & -- & -- & --\\
25 & -- &  -- &  0 (stable) & -- &  -- & 0.07 & 0.25\\
26--30$^*$ & -- & -- & -- & -- & -- & -- & --\\
31 & -- &  -- &  0 (stable) & -- &  -- & 0.08 & 0.25\\
32--33$^*$ & -- & -- & -- & -- & -- & -- & --\\
34 & 2 & 28.0 & 2.54$\times$10$^{-3}$ & 0.95$\times$10$^{-2}$ & 0.59 & --0.30 & 0.18\\
35 & 2 & 16.5 & 2.01$\times$10$^{-3}$ & 0.93$\times$10$^{-2}$ & 0.74 & --0.27 & 0.14\\
36 & -- &  -- & 0 (stable) & --  & -- & 0.10 & 0.17\\
37 & -- &  -- & 0 (stable) & --  & -- & 0.13 & 0.22\\
38$^*$ & -- & -- & -- & -- & -- & -- & --\\
39 & 2 & 18.0 & 0.18$\times$10$^{-3}$ & 1.71$\times$10$^{-2}$ & 15.13 & --0.02 & 0.16\\
40 & 1 & 24.0 & 2.99$\times$10$^{-3}$ & 1.57$\times$10$^{-2}$ & 0.84 & --0.31 & 0.22\\
41 & -- & --  & -- & -- &  -- & 0.05 & 0.24\\
42 & --  & -- & --  & -- & -- & 0.01 & 0.25\\
43 & 1 & 40.5 & 1.53$\times$10$^{-3}$ & 1.04$\times$10$^{-2}$ & 1.70 & --0.19 & 0.25\\
44 & 1 & 26.0 & 1.27$\times$10$^{-3}$ & 1.15$\times$10$^{-2}$ & 2.25 & --0.19 & 0.24\\
45 & 1 & 29.0 & 0.34$\times$10$^{-3}$ & 1.93$\times$10$^{-2}$ & 5.41 & --0.08 & 0.18\\
46, 48$^*$ & -- & -- & -- & -- & -- & -- & --\\
49 & 1 & 42.0 & 1.56$\times$10$^{-3}$ & 1.21$\times$10$^{-2}$ & 1.46 & --0.22 & 0.25\\
\hline
\end{tabular}
\begin{tablenotes}
    \item \hspace{86pt} {\small * Ri$_{\textrm{min}}$$>$0.30, no calculation was conducted.}
\end{tablenotes}
\end{table}

\clearpage
\begin{table}
\renewcommand\arraystretch{1.5}
\caption{Same as Table 2 but for the Celtic Sea.}
\centering
\begin{tabular}{c c c c c c c c}
\hline
Hour  & Mode  & $\lambda$ (m)  & $kc_i$ (s$^{-1}$)  & $N$ (s$^{-1}$) & $\tau/T_b$ & $\Phi_c$ & Ri$_c$\\
\hline
0 & 1 & 43.0 & 1.09$\times$10$^{-3}$ & 0.91$\times$10$^{-2}$ & 1.33 & --0.16 & 0.25\\
1 & -- &  -- & 0 (stable)  & --  & --  & 0.28 & 0.15\\
2 & -- &  -- & 0 (stable)  & --  & --  & 0.02 & 0.21\\
3 & -- &  -- & 0 (stable)  & --  & --  & 0.10 & 0.17\\
5--6$^*$ & -- & -- & -- & -- & -- & -- & --\\
7 & -- &  -- & 0 (stable)  & --  & --  & 0.10 & 0.19\\
8--11$^*$ & -- & -- & -- & -- & -- & -- & --\\
12 & -- &  -- & 0 (stable)  & --  & --  & 0.06 & 0.12\\
13 & 1 & 42.0 & 0.40$\times$10$^{-3}$ & 0.93$\times$10$^{-2}$ & 3.68 & --0.07 & 0.23\\
14$^*$ & -- & -- & -- & -- & -- & -- & --\\
15 & 2 & 28.0 & 0.27$\times$10$^{-3}$ & 0.32$\times$10$^{-2}$ & 1.85 & --0.10 & 0.21\\
16 & -- &  -- & 0 (stable)  & --  & --  & 0.38 & 0.12\\
17$^*$ & -- & -- & -- & -- & -- & -- & --\\
18 & 2 & 26.0 & 0.73$\times$10$^{-3}$ & 0.86$\times$10$^{-2}$ & 1.89 & --0.09 & 0.23\\
19 & -- &  -- & 0 (stable)  & --  & --  & 0.05 & 0.09\\
20 & -- &  -- & 0 (stable)  & --  & --  & 0.01 & 0.22\\
21 & -- &  -- & 0 (stable)  & --  & --  & 0.32 & 0.20\\
22--23$^*$ & -- & -- & -- & -- & -- & -- & --\\
\hline
\end{tabular}
\begin{tablenotes}
    \item \hspace{86pt} {\small * Ri$_{\textrm{min}}$$>$0.30, no calculation was conducted.}
\end{tablenotes}
\end{table}

\clearpage
\begin{table}
\renewcommand\arraystretch{1.5}
\caption{Values characterising the three seas: mean minimum gradient Richardson
number, and numbers and proportions of 1-hr records lying in various ranges of Ri. The
uncertainty of the estimates Ri$_{\textrm{c}}$ is about $\pm0.02$.}
\centering
\begin{tabular}{l c c c}
\hline
Parameter & Clyde Sea & Irish Sea & Celtic Sea\\
\hline
mean Ri$_{\textrm{min}}$ & 0.097 & 0.282 & 0.334\\
Number of 1-hr records: & & &\\
\hspace{0.60cm} total examined & 24 & 49 & 23\\
\hspace{0.60cm} with Ri$_{\textrm{min}}\ge\sfrac{1}{4}$ (stable)& 2 & 29 & 11 \\
\hspace{0.60cm} with Ri$_{\textrm{min}}$ $<$ $\sfrac{1}{4}$ (possibly unstable)& 22 & 20 & 12 \\
\hspace{0.60cm} with Ri$_{\textrm{min}}$ $<$ $\sfrac{1}{4}$, but stable (Ri$_{\textrm{min}}$ $\ge$ Ri$_{\textrm{c}}$) & 7 & 5 & 8 \\
\hspace{0.60cm} dynamically unstable (Ri$_{\textrm{min}}$ $<$ Ri$_{\textrm{c}}$) & 15 & 15 & 4\\
Proportion of 1-hr records: & & &\\
\hspace{0.60cm} stable & 38\% & 70\% & 83\%\\
\hspace{0.60cm} stable with Ri$_{\textrm{min}}$ $<$ $\sfrac{1}{4}$ & 29\%& 10\% & 35\%\\
\hspace{0.60cm} unstable (Ri$_{\textrm{min}}$ $<$ Ri$_{\textrm{c}}$) & 63\% & 30\% & 17\%\\
Marginal flows: & & &\\
(with 0.8 $<$ Ri$_{\textrm{min}}$$/$Ri$_{\textrm{c}}$ $<$ 1.2, i.e., --0.105 $<$ $\Phi_{\textrm{c}}$ $<$ 0.095) & & &\\
\hspace{0.60cm} number of marginal flows & 10 & 12 & 7\\
\hspace{0.60cm} proportion of marginal flows & 42\% & 24\% & 30\%\\
Very stable flows: & & &\\
\hspace{0.60cm} number with Ri$_{\textrm{min}}$$/$Ri$_{\textrm{c}}$ $>$ 1.2 & 4 & 26 & 15\\
\hspace{0.60cm} proportion with Ri$_{\textrm{min}}$$/$Ri$_{\textrm{c}}$ $>$ 1.2 & 17\% & 53\% & 65\%\\
\hline
\end{tabular}

\end{table}

%

\begin{figure}[!t]
\begin{center}
\noindent\includegraphics[width=16cm]{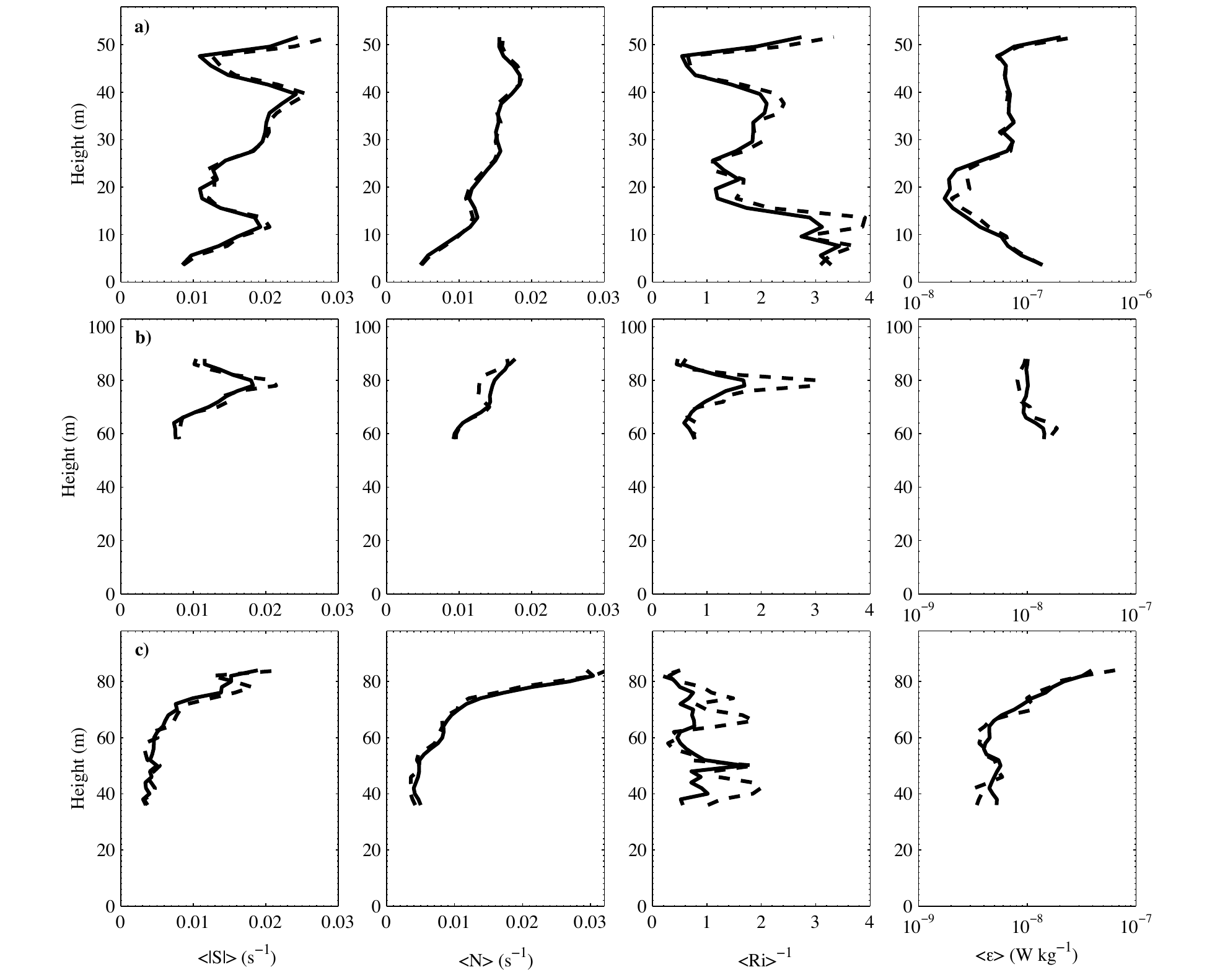}
\end{center}
\caption{Mean profiles for the three seas. (a) The Clyde, (b) the Irish and (c) the Celtic Sea.
Left to right: the modulus of the mean shear, $<$$|S|$$>$; the modulus of the buoyancy frequency,
$<$$N$$>$; the inverse mean Ri where $<$Ri$>$ $=$ $<$$N^2$$>$$/$$<$$|S|^2$$>$; and the rate of dissipation of turbulent kinetic energy per unit mass, $<$$\varepsilon$$>$. Solid lines indicate the means of all the data, and the
dashed lines indicate data in unstable conditions (Ri$_{\textrm{min}}$$<$ Ri$_{\textrm{c}}$).}\label{fig01}
\end{figure}

\begin{figure}[!t]
\begin{center}
\noindent\includegraphics[width=16cm]{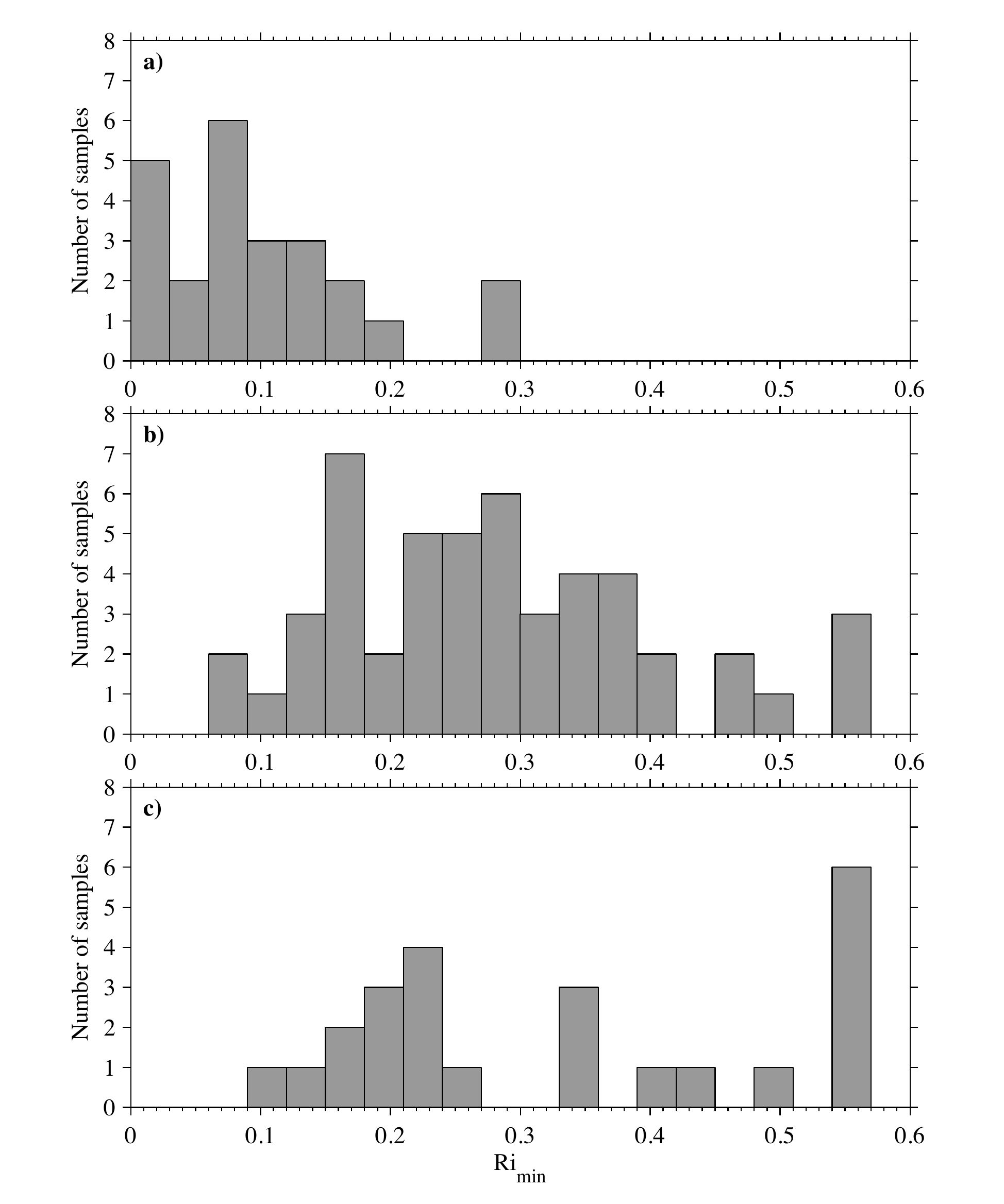}
\end{center}
\caption{Histogram of Ri$_{\textrm{min}}$ showing values for each of the three seas, (a) The Clyde,
(b) the Irish and (c) the Celtic Sea.}\label{fig02}
\end{figure}

\begin{figure}[!t]
\begin{center}
\noindent\includegraphics[width=16cm]{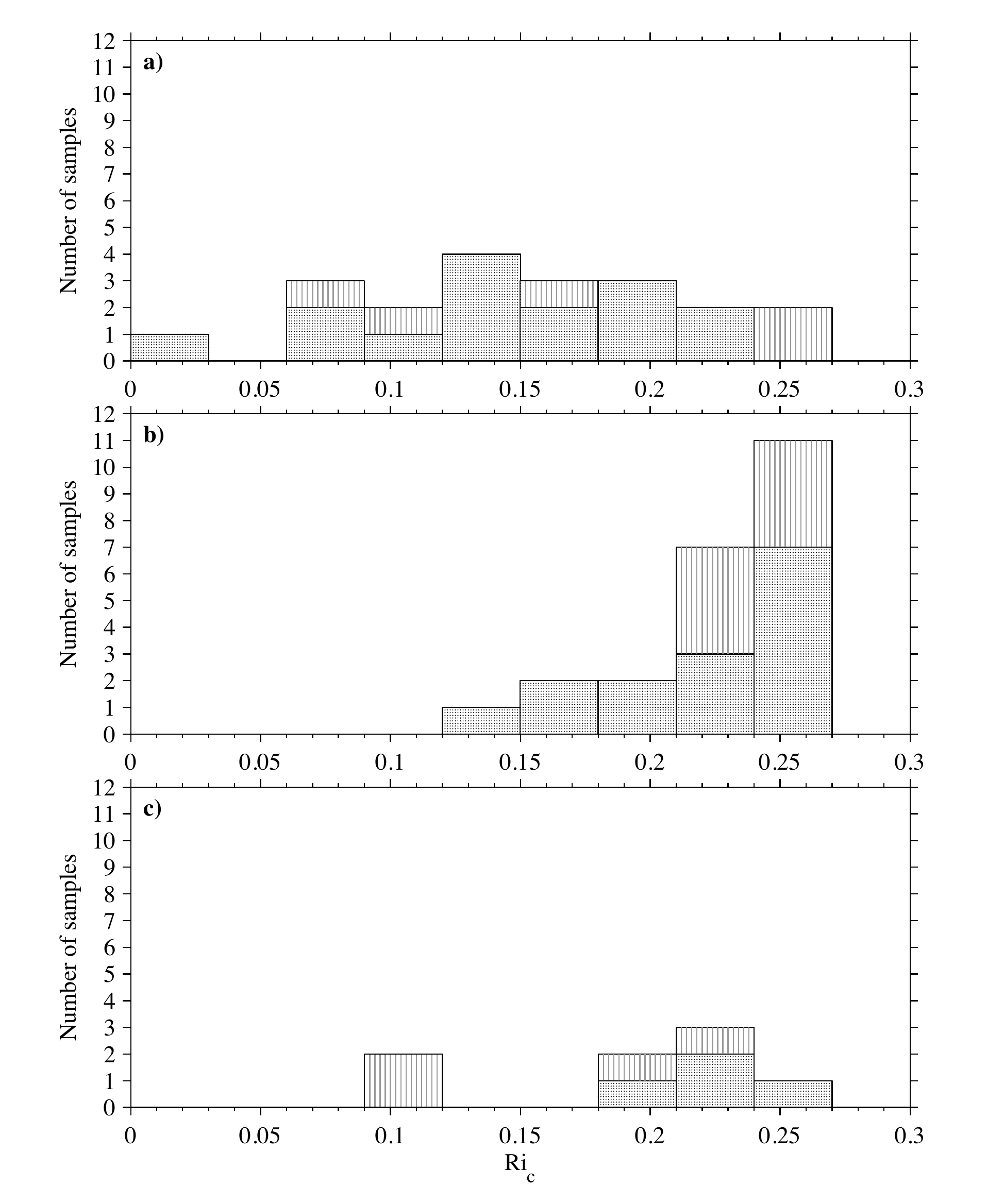}
\end{center}
\caption{Histogram of Ri$_{\textrm{c}}$ for all the flows in which Ri$_{\textrm{min}}/$Ri$_{\textrm{c}}$ $<$ 1.2 for each of the three seas, (a) The Clyde, (b) the Irish and (c) the Celtic Sea. The unstable flows with Ri$_{\textrm{min}}$$<$ Ri$_{\textrm{c}}$ are stippled and the stable flows with Ri$_{\textrm{min}}\ge$ Ri$_{\textrm{c}}$ are hatched.}\label{fig03}
\end{figure}

\begin{figure}[!t]
\begin{center}
\noindent\includegraphics[width=16cm]{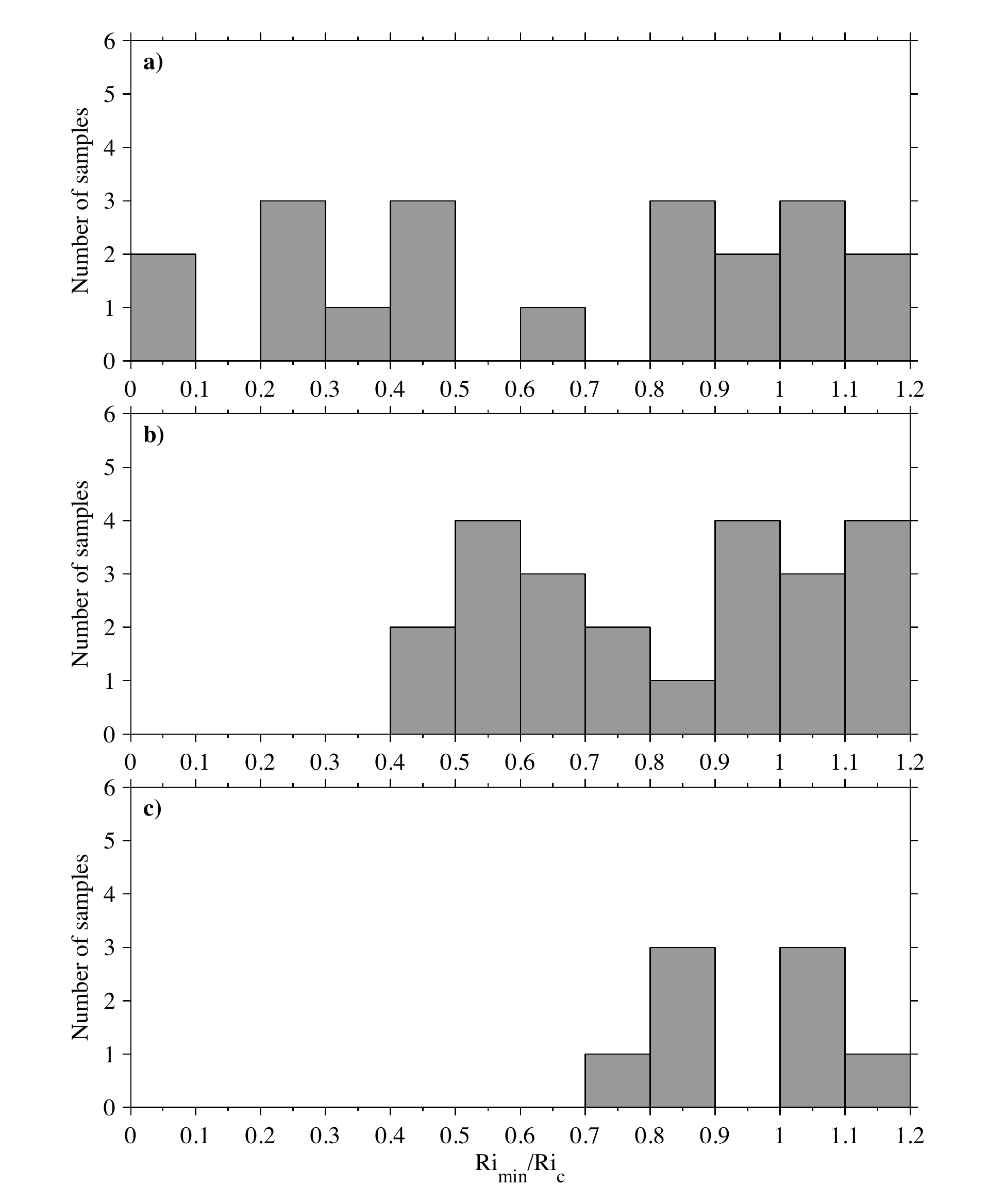}
\end{center}
\caption{Histogram of Ri$_{\textrm{min}}/$Ri$_{\textrm{c}}$ (= $(1 + \Phi_{\textrm{c}})^2$) $<$ 1.2, showing values for each of the three seas, (a) The Clyde, (b) the Irish and (c) the Celtic Sea. The number of values of
Rimin/Ric exceeding 1.2 is given in Table 4.}\label{fig04}
\end{figure}

\begin{figure}[!t]
\begin{center}
\noindent\includegraphics[width=16cm]{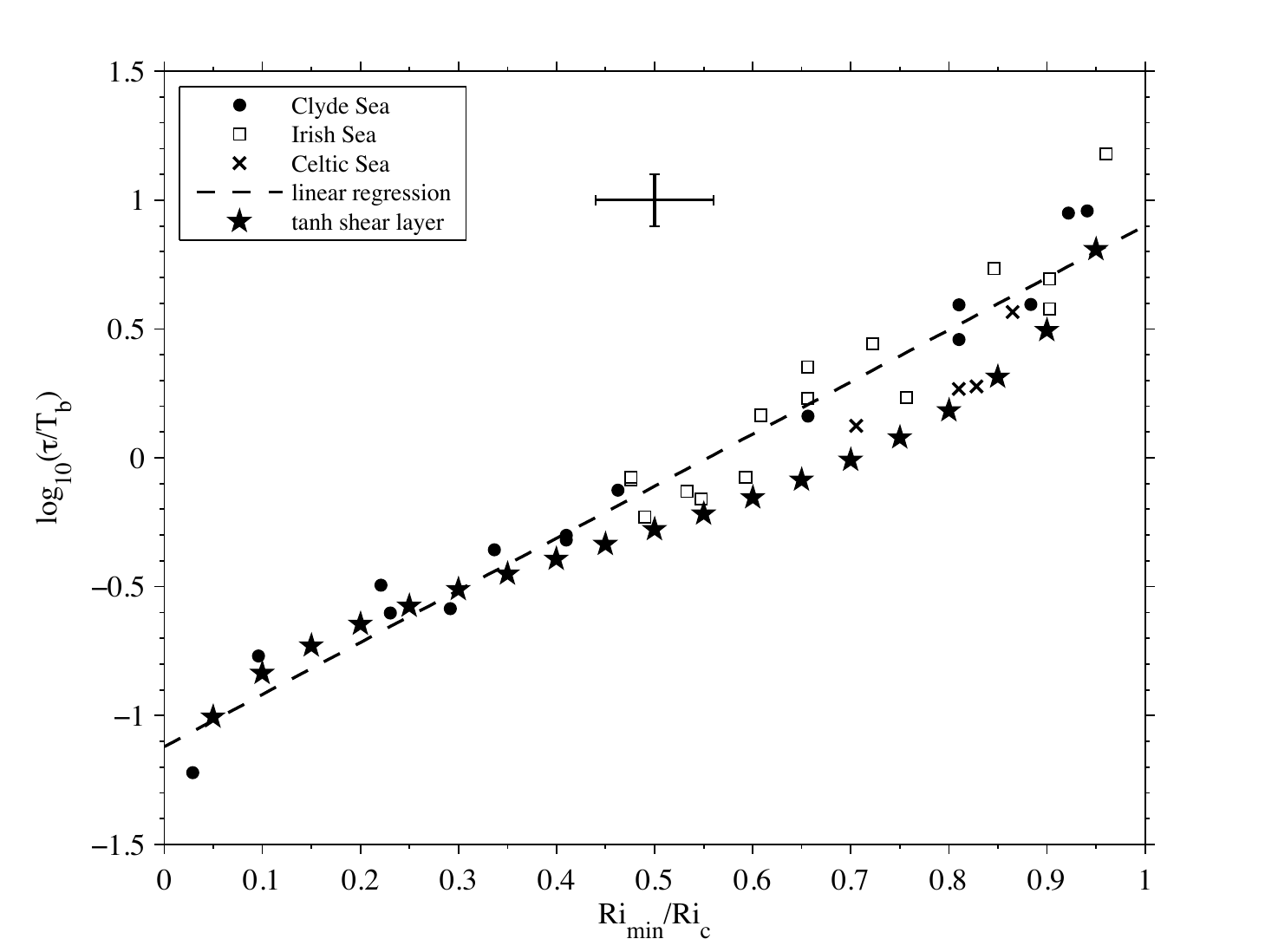}
\end{center}
\caption{The variation of log$_{10}(\tau/T_b)$ with Ri$_{\textrm{min}}/$Ri$_{\textrm{c}}$ (= $(1 + \Phi_{\textrm{c}})^2$). Data from the three seas are shown as: $\bullet$ Clyde, $\square$ Irish and $\mathbf{\times}$ Celtic. The cross indicates uncertainty of the estimated values. The uncertainty in Ri$_{\textrm{min}}/$Ri$_{\textrm{c}}$ (about 0.06) shown in the figure is estimated from the uncertainties in estimates of Ri due to the separation of the ADCP
and FLY data, judged to be about 0.03 as for Ri$_{\textrm{min}}$ in the Appendix. (Since Ri$_{\textrm{min}}/$Ri$_{\textrm{c}}$ = $(1 + \Phi_{\textrm{c}})^2$, an estimated uncertainty of $\Delta\Phi_{\textrm{c}}$ $\approx$ 0.02 in $\Phi_{\textrm{c}}$ gives an uncertainty in Ri$_{\textrm{min}}/$Ri$_{\textrm{c}}$ of about $2(1+\Phi_{\textrm{c}})\Delta\Phi_{\textrm{c}}$ $<$ 0.08, since $|\Phi_{\textrm{c}}|$ is generally $<$ 1, reasonably
consistent with the above estimate.) Data for a simple hyperbolic tangent shear layer is shown as $\bigstar$. }\label{fig05}
\end{figure}

\begin{figure}[!t]
\begin{center}
\noindent\includegraphics[width=16cm]{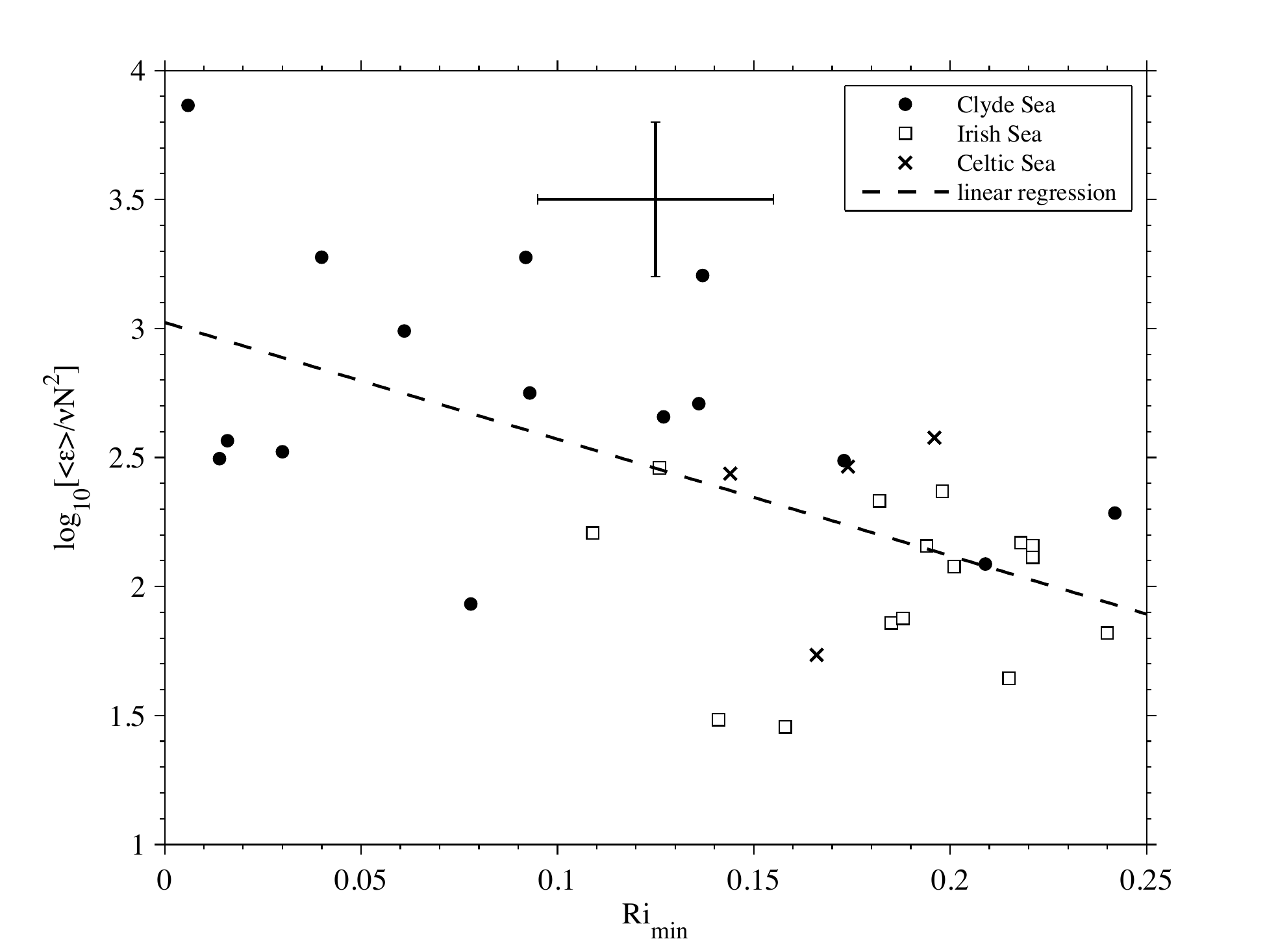}
\end{center}
\caption{The variation of log$_{10}($$<$$\varepsilon$$>$$/\nu N^2$) with Ri$_{\textrm{min}}$. Data from the three seas are shown as: $\bullet$ Clyde, $\square$ Irish and $\mathbf{\times}$ Celtic. The dashed line, log$_{10}($$<$$\varepsilon$$>$$/\nu N^2$) = --4.52 Ri$_{\textrm{min}}$ + 3.02, indicates a regression line with a correlation coefficient of --0.59, with the 95\% confidence interval being (--0.77 --0.32), i.e. statistically significant. The cross
indicates uncertainty of the estimated values. The error in log$_{10}($$<$$\varepsilon$$>$$/\nu N^2$) derives
largely from the uncertainty in estimates of about a factor of 2 in $\varepsilon$, and those in Ri$_{\textrm{min}}$ from the horizontal separation of the ADCP and FLY sites (see the Appendix).}\label{fig06}
\end{figure}

\begin{figure}[!t]
\begin{center}
\noindent\includegraphics[width=16cm]{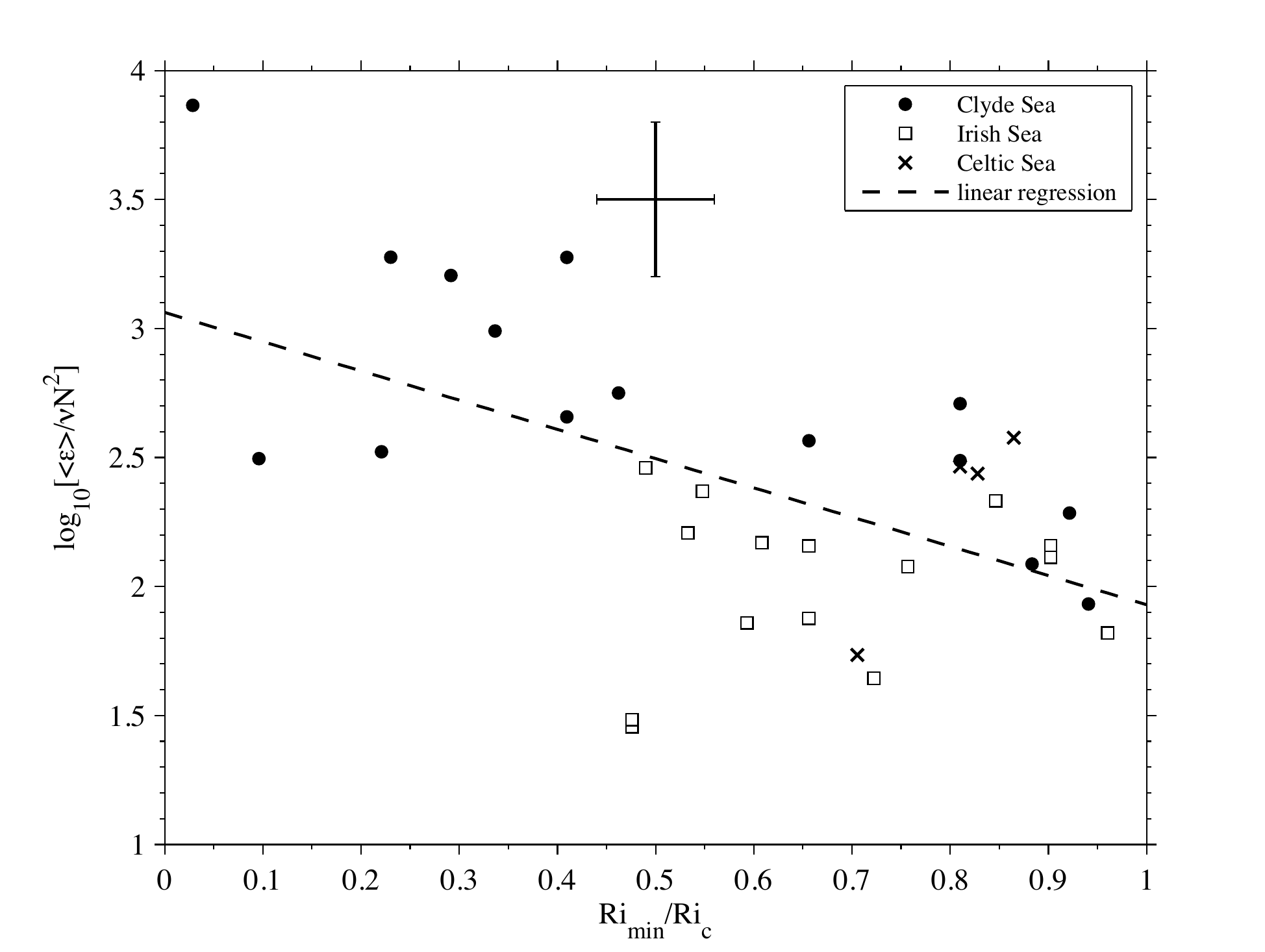}
\end{center}
\caption{The variation of log$_{10}($$<$$\varepsilon$$>$$/\nu N^2$) with Ri$_{\textrm{min}}/$Ri$_{\textrm{c}}$ = $(1 + \Phi_{\textrm{c}})^2$. Data from the three seas are shown as: $\bullet$ Clyde, $\square$ Irish and $\mathbf{\times}$ Celtic. The dashed line, log$_{10}($$<$$\varepsilon$$>$$/\nu N^2$) = --1.13 Ri$_{\textrm{min}}/$Ri$_{\textrm{c}}$ + 3.05, indicates a regression line with a correlation coefficient --0.54, with the 95\% confidence interval being (--0.74 --0.25), i.e. statistically significant. The
cross indicates uncertainty of the estimated values. See caption of Fig. 5.}\label{fig07}
\end{figure}

\begin{figure}[!t]
\begin{center}
\noindent\includegraphics[width=16cm]{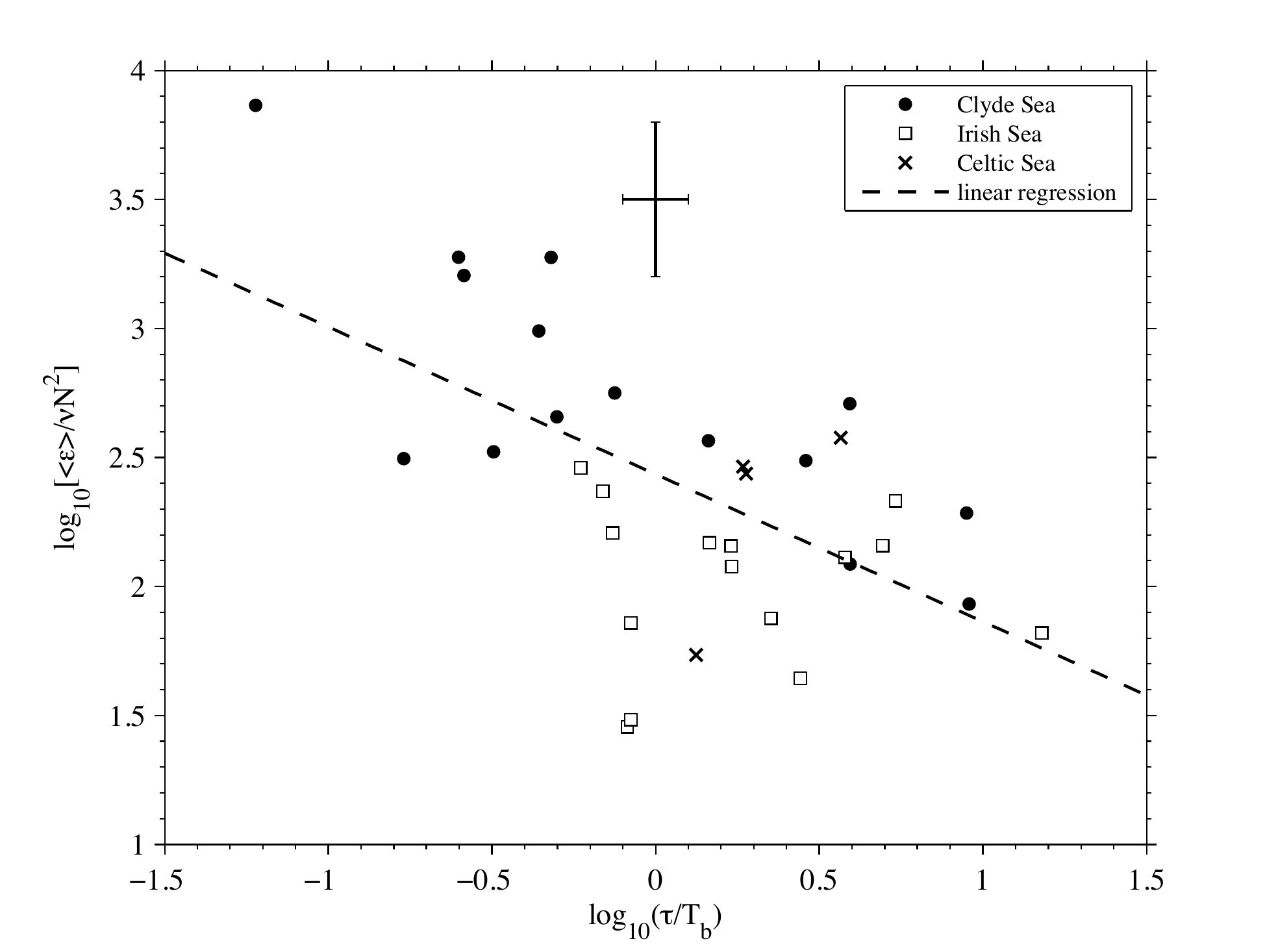}
\end{center}
\caption{The variation of log$_{10}($$<$$\varepsilon$$>$$/\nu N^2$) with log$_{10}(\tau/T_b)$. Data from the three seas are shown as: $\bullet$ Clyde, $\square$ Irish and $\mathbf{\times}$ Celtic. The dashed line, log$_{10}($$<$$\varepsilon$$>$$/\nu N^2$) = --0.57 log$_{10}(\tau/T_b)$ + 2.44, indicates a regression line with a correlation coefficient --0.57, with the 95\% confidence interval being (--0.76 --0.29), i.e. statistically significant. The
cross indicates uncertainty of the estimated values. See caption of Fig. 6.}\label{fig08}
\end{figure}

\begin{figure}[!t]
\begin{center}
\noindent\includegraphics[width=16cm]{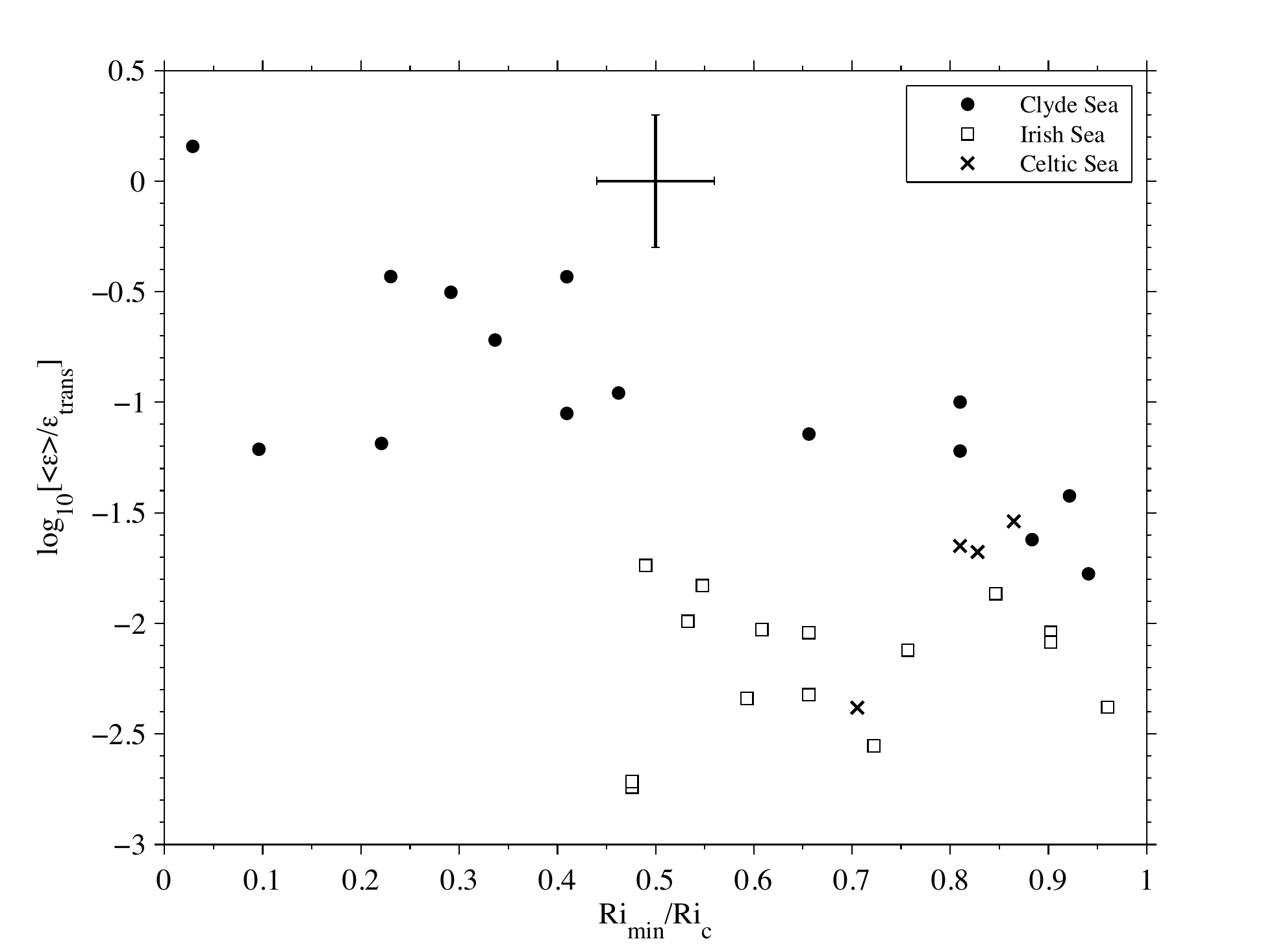}
\end{center}
\caption{The variation of log$_{10}($$<$$\varepsilon$$>$$/\varepsilon_{trans}$) with Ri$_{\textrm{min}}/$Ri$_{\textrm{c}}$, where $\varepsilon_{trans}$ denotes the dissipation rate at the W--T boundary as given by (1). The cross indicates uncertainty of the estimated values.}\label{fig09}
\end{figure}

\begin{figure}[!t]
\begin{center}
\noindent\includegraphics[width=16cm]{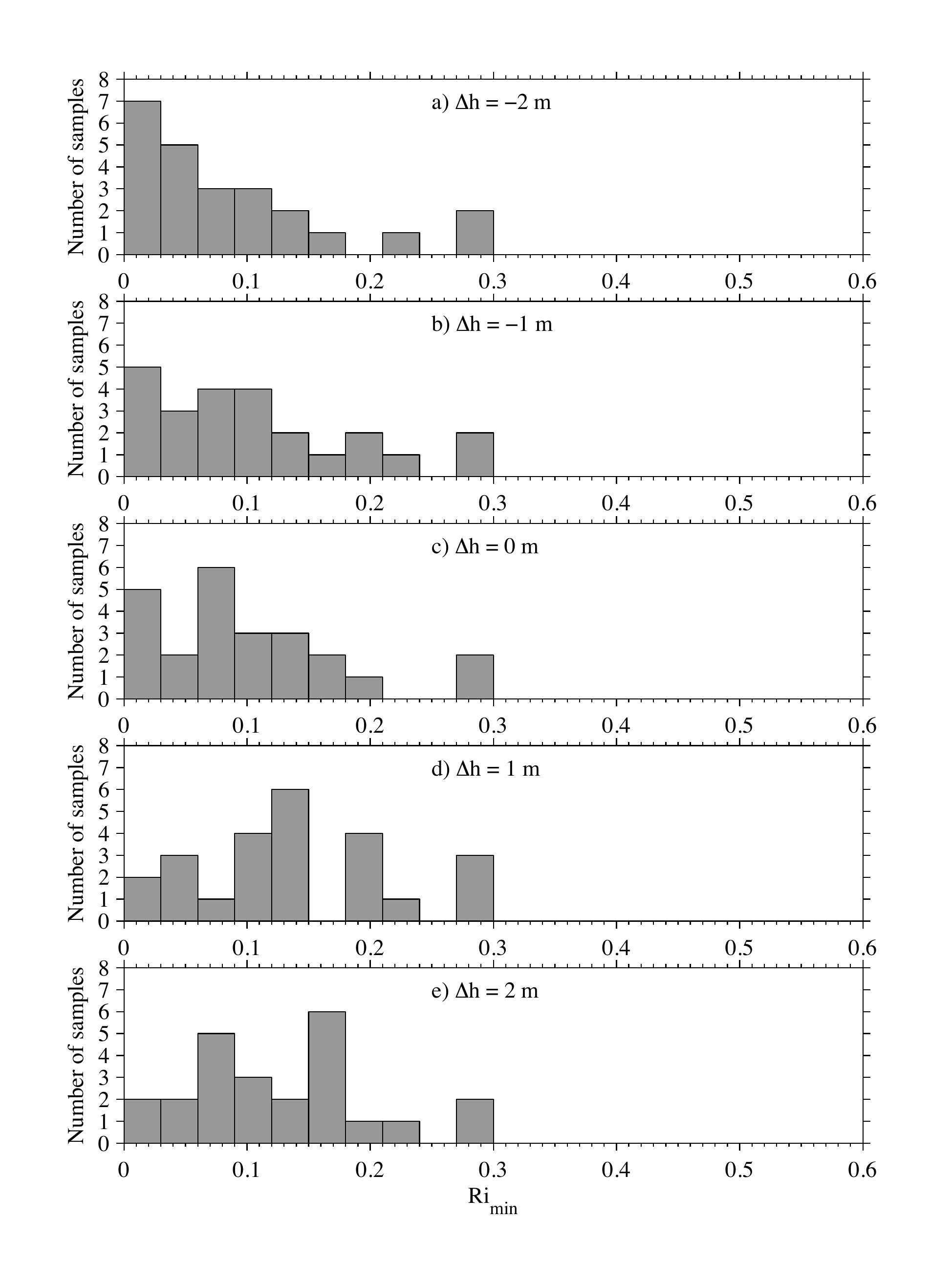}
\end{center}
\caption{The distributions of Ri$_{\textrm{min}}$ in the Clyde Sea at (top to bottom) $\Delta h = -2$ m to
+2 m.}\label{figA1}
\end{figure}

\end{document}